\documentstyle[aps,multicol]{revtex}
\renewcommand{\narrowtext}{\begin{multicols}{2} \global\columnwidth20.5pc}
\renewcommand{\widetext}{\end{multicols} \global\columnwidth42.5pc}
\input{epsf.tex}

\def\inseps#1#2{\def\epsfsize##1##2{#2##1} \centerline{\epsfbox{#1}}}

\begin{document}
\draft

\title{Statistical properties of the low-temperature conductance
peak-heights for Corbino discs in the quantum Hall regime
}

\author{N. R. Cooper$^1$, B. I. Halperin$^2$, Chin-Kun Hu$^3$, and I. M. Ruzin$^4$}
\address{$^1$Institut
Laue-Langevin, B.P. 156, 38042 Grenoble, Cedex 9, France. \\
$^2$Department of Physics, Harvard University, Cambridge, MA 02138,
USA.\\ $^3$Institute of Physics, Academia Sinica, Nankang, Taipei
11529, Taiwan.\\ $^4$Department of Molecular and Microbiology, Tufts
University, Boston, MA 02111, USA.}

\date{18 October 1996}

\maketitle

\begin{abstract}

A recent theory has provided a possible explanation for the
``non-universal scaling'' of the low-temperature conductance (and
conductivity) peak-heights of two-dimensional electron systems in the
integer and fractional quantum Hall regimes.  This explanation is
based on the hypothesis that samples which show this behavior contain
density inhomogeneities.  Theory then relates the non-universal
conductance peak-heights to the ``number of alternating percolation
clusters'' of a continuum percolation model defined on the
spatially-varying local carrier density.  We discuss the statistical
properties of the number of alternating percolation clusters for
Corbino disc samples characterized by random density fluctuations
which have a correlation length small compared to the sample size.
This allows a determination of the statistical properties of the
low-temperature conductance peak-heights of such samples.  We focus on
a range of filling fraction at the center of the plateau transition
for which the percolation model may be considered to be critical.  We
appeal to conformal invariance of critical percolation and argue that
the properties of interest are directly related to the corresponding
quantities calculated numerically for bond-percolation on a cylinder.
Our results allow a lower bound to be placed on the non-universal
conductance peak-heights, and we compare these results with recent
experimental measurements.

\end{abstract}



\narrowtext

\section{Introduction}
\label{sec:introduction}

Experimental studies of the dissipative conductivity $\sigma_{xx}$ of
two-dimensional electron systems in the integer and fractional quantum
Hall regimes have shown this to obey an unusual ``non-universal
scaling'' at very low temperatures\cite{nonuniexpt}.
``Scaling'' refers to the observation that the height of the peak in
$\sigma_{xx}$ associated with a transition between a plateau with
quantized Hall conductivity $\sigma_1$ and one with quantized Hall
conductivity $\sigma_2$ is proportional to $|\sigma_1-\sigma_2|$ for
all well-developed peaks in a given sample.  The constant of
proportionality is found to fluctuate between samples, so this scaling
is described as ``non-universal''.  This behavior is in contrast to
theoretical predictions of scaling with a {\it universal} prefactor of
$0.5$\cite{Kucera,huo,ruzin}.
Non-universal scaling was most clearly demonstrated in recent
experiments\cite{rokhinsongoldman} on two Corbino disc samples for
which prefactors of approximately $0.2$ and $0.3$ were found.

In Ref.~\onlinecite{rch} an explanation for non-universal scaling was
proposed. This explanation is based on the hypothesis that in samples
exhibiting this phenomenon there exist fluctuations in the electron
density on scales large compared to microscopic length scales.  From a
classical calculation of the transport through the resulting
inhomogeneous conductor, it was shown that, in samples with such
inhomogeneities, non-universal scaling appears for all plateau
transitions that have reached their low-temperature limiting forms (we
shall refer to such transitions as ``well-developed'').  We will
concentrate on the conclusions that were obtained for Corbino disc
samples (analogous results hold for Hall bars\cite{rch}).  For a
well-developed transition between two quantized Hall plateaus with
Hall conductivities $\sigma_1$ and $\sigma_2$ the two-terminal {\it
conductance} of a Corbino disc sample takes the form\cite{rch}
\begin{equation}
\label{eq:lowtg}
G = M |\sigma_1-\sigma_2|,
\end{equation}
where $M$ is an {\it integer} related to the geometrical properties of
a classical percolation problem defined on the spatially-varying
filling fraction $\nu(\bbox{r})$ [the local filling fraction
$\nu(\bbox{r})$ is defined as the ratio of the local electron number
density $n(\bbox{r})$ to the density of flux quanta $eB/h$]. We refer
to $M$ as the ``number of alternating percolation clusters''; this
quantity will be defined below, and the reason for this name will then
become clear.  As the magnetic field is varied, such that the sample
sweeps through the plateau transition, the number of alternating
percolation clusters passes through a sequence of integer values,
starting and finishing with $M=0$ (in the plateau regions).
Significantly, although the specific sequence of values that $M$
passes through will in general differ for different configurations of
the density fluctuations, the {\it same} sequence is predicted to
occur for all well-developed transitions within a given sample.  In
particular, the maximum value $M^{max}$ will be the same for all such
transitions.  Thus, within the theory of Ref.~\onlinecite{rch}, the
maximum heights of all well-developed {\it conductance} peaks of a
Corbino disc sample show a scaling with a sample-dependent, but
integer, prefactor $M^{max}$.  If the measurements of the conductance
$G$ are used to define an observed conductivity $\sigma_{xx}$ via the
usual relation $\sigma_{xx}\equiv A_0 G$ where $A_0\equiv
(1/2\pi)\ln(r_2/r_1)$ is the geometrical aspect ratio of the Corbino
disc determined by its inner $r_1$ and outer $r_2$ radii, then the
peak-heights of the conductivity will also exhibit non-universal
scaling.  In this case the non-universal prefactor is $A_0 M^{max}$,
which can vary between samples due to both fluctuations in the integer
$M^{max}$ and variations in the aspect ratio $A_0$. In
Ref.~\onlinecite{rch} it was shown that the non-universal prefactors
for the conductivity peak-heights of the two Corbino discs studied in
Ref.~\onlinecite{rokhinsongoldman} are both consistent with the form
$A_0 M^{max}$ if $M^{max}=1$ in each case.

It is of interest to understand the statistical properties of the
integer $M^{max}$ which determines the non-universal scaling
prefactor.  This is the issue that we address in the present paper.
We study the probability distribution of the number of alternating
percolation quantities $M$ for a Corbino disc sample as a function of
the aspect ratio $A_0$, denoting the probabilities $P_M(A_0)$.  As
discussed in Section~\ref{sec:continuum}, we study these probabilities
for an ensemble of Corbino disc samples characterized by density
fluctuations with a scale much less than the sample dimensions, and
within a narrow range of filling fraction at the center of the quantum
Hall transition, in which case the percolation model determining $M$
can be considered to be critical.  By appealing to conformal
invariance of critical percolation, we argue that the analogous
probabilities which have recently been calculated numerically for
bond-percolation in the cylindrical geometry\cite{huhalperin} can be
directly related to $P_M(A_0)$.  For some disorder configurations, the
maximum value $M^{max}$ could lie outside the region of filling
fraction that we address, so our results provide information on a {\it
lower bound} to $M^{max}$.  We show that, if the two samples studied
in Ref.~\onlinecite{rokhinsongoldman} are drawn from the statistical
ensemble that we have assumed, the combined probability for
$M^{max}=1$ in both of these samples is less than $15\%$.  We view
this small probability as evidence that short-range isotropic density
inhomogeneities may not well represent the experimental samples, and
suggest that density fluctuations on a scale comparable to the sample
size could be present.

\section{Relation to Continuum Percolation}
\label{sec:continuum}

\begin{figure}
\vskip-5mm
\inseps{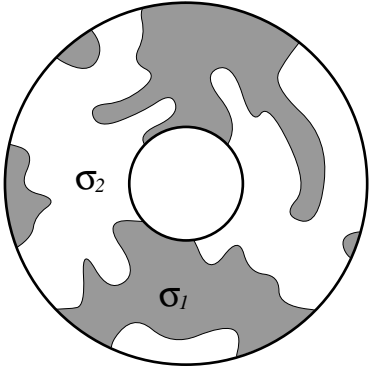}{0.28}
\caption{Schematic diagram of the distribution of local conductivity
in a Corbino disc at very low temperature, in the transition region
between two quantized Hall plateaus. Black and white colors represent
regions where the local conductivity is quantized, with Hall
conductivities $\sigma_1$ and $\sigma_2$ respectively.}
\label{fig:saturate}
\end{figure}

We begin by explaining how the integer $M$ appearing in
Eq.~(\ref{eq:lowtg}) is related to the geometrical properties of a
continuum percolation model (the reader is referred to
Ref.~\onlinecite{rch} for a detailed discussion). The connection to
percolation arises when one considers the form of the spatial
distribution of the local conductivity in the sample.  In the theory
of Ref.~\onlinecite{rch}, the conductivity tensor at a position
$\bbox{r}$ is assumed to be specified by the local value of the
filling fraction $\nu(\bbox{r})$, and equal to the conductivity of a
homogeneous system with this filling fraction.  The inhomogeneous
samples are assumed to contain short-range disorder on length scales
much smaller than those of the density inhomogeneities, so the
appropriate homogeneous system to consider is one which retains this
impurity scattering. Consequently, the dependence of the local
conductivity on the local filling fraction $\sigma(\nu)$ exhibits a
quantized Hall effect, with the conductivity tensor taking quantized
values over finite ranges of filling fraction.  At low temperatures
(which we consider in detail below), the transition regions between
quantized Hall plateaus are very narrow in filling fraction and the
conductivity is quantized at almost all values of the filling
fraction.  These quantized plateau regions lead to spatially-extended
regions in the inhomogeneous sample within which the local
conductivity takes the same, quantized value (in these regions the
local filling fraction remains within a quantized Hall plateau). In
particular, consider a situation in which the average filling fraction
of the inhomogeneous sample $\overline\nu$ is close to the narrow
range of filling fractions within which the homogeneous sample
displays a transition between plateaus with quantized Hall
conductivities $\sigma_1$ and $\sigma_2$.  We denote the filling
fraction at which the quantum Hall transition of the homogeneous
sample is centered by $\nu_c$, and its width in filling fraction by
$\delta\nu$.  In the inhomogeneous sample, regions in which the local
filling fraction is less than $\nu_c-\delta\nu/2$ will have a local
conductivity that is quantized with Hall conductivity $\sigma_1$ (we
shall refer to these as ``$\sigma_1$-regions''), and regions where the
local filling fraction is larger than $\nu_c+\delta\nu/2$ will have a
local conductivity tensor that is quantized according to $\sigma_2$
(``$\sigma_2$-regions'').  As temperature decreases, the regions of
intermediate filling fraction that spatially separate the $\sigma_1$-
and $\sigma_2$-regions, and inside of which the local conductivity
tensor is not quantized, become progressively narrower, due to the
progressive decrease in the width $\delta\nu$ of the plateau
transition for an infinite homogeneous sample.  In the limit of low
temperatures, for which the plateau transition of the finite
inhomogeneous sample is ``well-developed'' and Eq.~(\ref{eq:lowtg})
applies, these intermediate regions may be considered to be
infinitesimally narrow.  The sample then divides cleanly into
$\sigma_1$- and $\sigma_2$-regions separated by sharp boundaries along
lines on which the filling fraction has the threshold value $\nu_c$
(this is the filling fraction at which the plateau transition occurs
for an infinite homogeneous sample in the limit of zero temperature).
The spatial distribution of the conductivity is determined by a
continuum percolation model defined on the spatially-varying filling
fraction $\nu(\bbox{r})$ with the threshold $\nu_c$: in a region where
the filling fraction is below (above) $\nu_c$ the local conductivity
is quantized with a Hall conductivity $\sigma_1$ ($\sigma_2$).
Figure~\ref{fig:saturate} illustrates a particular configuration of
the local conductivity, for an average filling fraction close to
$\nu_c$ such that the two regions occupy almost equal area.

As discussed in detail in Ref.~\onlinecite{rch}, in the limit of low
temperatures when the {\it local} conductivity tensor can be assumed
to be purely off-diagonal, dissipation does not occur in the bulk of
the sample but only at junctions formed between the $\sigma_1$- and
$\sigma_2$-clusters and the contacts.  The net conductance of the
sample depends only on the {\it topology} of the conductivity
distribution.  Specifically, the two-terminal conductance of the
sample is found to be given by Eq.~(\ref{eq:lowtg}), with the ``number
of alternating percolation clusters'' $M$ defined by
\begin{equation}
\label{eq:m}
M \equiv min(n_1, n_2) ,
\end{equation}
where $n_i$ is the number of clusters of type $\sigma_i$ that connect
between inner and outer contacts of the Corbino disc.  The
configuration illustrated in Figure~\ref{fig:saturate} is a case in
which two $\sigma_1$-clusters and two $\sigma_2$-clusters connect the
contacts, so $n_1=n_2=2$ and $M=2$.

We will briefly discuss the evolution of $M$ as the sample is swept
through a transition between the $\sigma_1$ and $\sigma_2$ quantum
Hall plateaus.  Consider first a situation in which the average
filling fraction $\bar\nu$ is sufficiently below $\nu_c$ that the
sample is in the $\sigma_1$ quantized Hall plateau.  The conductivity
at almost all points in the sample is quantized at $\sigma_1$ so the
inner and outer contacts are connected by a {\it single} cluster of
this kind. Therefore $n_1=1$ and $n_2=0$, and, by Eqs.~(\ref{eq:m})
and~(\ref{eq:lowtg}), $M=0$ and $G=0$. This correctly attributes a
vanishingly small conductance to the sample in this quantized Hall
plateau.  When $\bar\nu$ is sufficiently larger than $\nu_c$, the
sample is in the $\sigma_2$ quantized Hall plateau, and a similar
picture emerges (with the roles of $\sigma_1$ and $\sigma_2$
reversed): $n_2=1$ and $n_1=0$ and the conductance vanishes $G=0$.

As the average filling fraction is swept through $\nu_c$, such that
the sample passes between the two quantized Hall plateaus, there
appear distributions of the two phases that are quite different from
those in the plateau regions.  It is straightforward to convince
oneself that the topology of the Corbino disc constrains the only
other possible values of the pair $(n_1,n_2)$ to the cases $(0,0),
(1,1), (2,2)\ldots$, for which $M=0,1,2\ldots$.  In the course of the
plateau transition, the distribution of the conductivity will pass
through one or more of these different topological configurations, so
$M$ will pass through a sequence of integer values and a step-like
peak in the conductance $G$ will appear [as noted in
Ref.~\onlinecite{rch}, it is possible that no peak appears, since the
sequence $(1,0) \rightarrow (0,0)
\rightarrow (0,1)$ is permitted]. 
The sequence of topological configurations and consequently the shape
and height of the conductance peak depend on the specific spatial
distribution of the filling fraction $\nu(\bbox{r})$ on which the
percolation model is defined. The statistical properties of the peak
heights and shapes therefore depend on the statistical form of the
density inhomogeneities in the sample.

We will assume that the spatial fluctuations in the filling fraction
$\nu(\bbox{r})$ are isotropic and homogeneous and have a correlation
length $R_c$ that is small compared to the sample size $L$.  The form
of the long-range density fluctuations in typical quantum Hall samples
is not known, so it is not clear if the form we study is generally
appropriate.  However, in the absence of any prior knowledge, it is
natural to consider the fluctuations to be homogeneous and isotropic.
Furthermore, provided the correlation length $R_c$ is small compared
to the sample size, universality of percolation causes the statistical
properties at large length scales to be insensitive to the details of
the density fluctuations at a scale $R_c$.  The large-scale properties
of the resulting percolation model are controlled by the correlation
length $\xi$, which diverges at the percolation threshold according
to\cite{isichenko,staufferbook}
\begin{equation}
\xi\simeq R_c
|\bar\nu-\bar{\nu}_c|^{-4/3}.
\end{equation}
(The percolation threshold of an infinite inhomogeneous sample
$\bar{\nu}_c$ is equal to the critical filling fraction of an infinite
homogeneous sample $\nu_c$ if the density fluctuations are
statistically symmetric about $\bar\nu$.)  In particular, the
transition region between the two quantized Hall plateaus occurs
within the range of average filling fraction $\bar\nu$ close to the
percolation threshold $\bar{\nu}_c$ for which the correlation length
$\xi$ becomes comparable to the system size, $\xi\gtrsim L$.  

We will restrict attention to the narrow range at the center of the
transition for which $\xi\gg L$.  In this central range, the
distribution of the two phases is equivalent to the distribution
exactly at the critical point $\bar\nu = \bar{\nu}_c$: we will refer
to this range of filling fraction as the ``critical region'' of the
transition.  In the following, we discuss how the properties in the
critical region of the continuum percolation model may be related to
numerical studies of a critical lattice model.

\section{Universality and Conformal Invariance of Critical Percolation}
\label{sec:universality}

It is well-established that there exist ``universal'' properties of
percolation models that are the same for all models of percolation
within a given ``universality class'' (specified by the symmetry and
dimensionality of the model).  In particular, it is believed that at
their respective critical points all percolation models within the
same universality class give rise to the same probability distribution
of configurations coarse-grained on a length scale $\Lambda$ in the
limit $\Lambda\gg a$, where $a$ is the largest microscopic length
scale of the model\cite{isichenko,staufferbook}. Any property that is
a function of this coarse-grained probability distribution will also
be a universal property of critical percolation.  We anticipate (as we
discuss further below) that the probabilities $P_M(A_0)$ in a finite
system are universal properties of two-dimensional isotropic
percolation at criticality, since these depend on the structure of
clusters on large scales (of order the system size $L$) and can be
expected to be insensitive to the introduction of coarse-graining over
small length scales $\Lambda\ll L$.

\begin{figure}
\vskip-5mm
\inseps{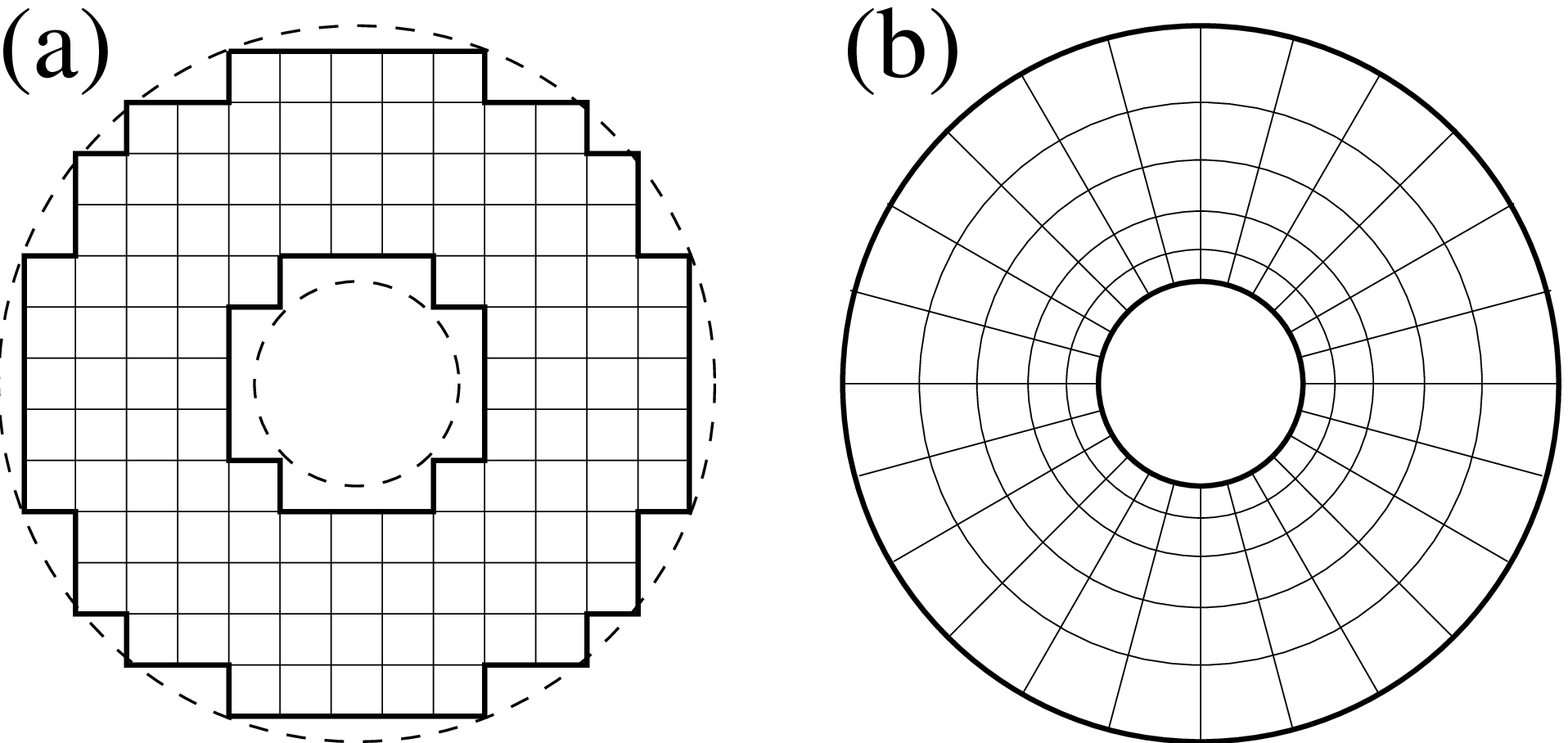}{0.4}
\vskip2mm
\caption{Two possible discretizations of the Corbino disc. (a) Uniform
square lattice with boundaries chosen to best reproduce the circular
contacts (dashed). (b) Conformal transformation of a square lattice
defined on a cylinder: the lattice constant varies continuously with
position, but the lattice is everywhere locally square. }
\label{fig:discretize}
\end{figure}

If this is the case, these probabilities may be determined by studying
bond-percolation on a lattice, which falls within the same
universality class as the continuum percolation in which we are
primarily interested.  One can achieve a discretization of the Corbino
disc that is suitable for studying bond-percolation by introducing a
finite square lattice of the form illustrated in
Fig.~\ref{fig:discretize}(a) (one could equally well use a triangular,
honeycomb or random lattice, all of which lead to isotropic behavior
on large length scales).  However, motivated by the fact that a
percolation model is defined by local rules, it is natural to extend
the concept of universality to include percolation models whose
properties vary slowly in space, provided that {\it locally} these
models are suitable representations of critical percolation (within
the appropriate universality class).  This hypothesis forms the basis
for the application of conformal field theory to percolation and other
critical systems\cite{cardyreview}. In particular, bond-percolation on
any {\it conformal transformation} of a square lattice presents a good
model of the critical percolation in which we are interested, since
this transformation generates a lattice that is {\it locally} square.
A discretization of the Corbino disc by a lattice of this kind is
illustrated in Fig.~\ref{fig:discretize}(b).

The work that we present in this paper and the conclusions that we
will draw are based on the hypothesis that a critical bond-percolation
model defined on the discretization of the Corbino disc illustrated in
Fig.~\ref{fig:discretize}(b) may be used to determine the
probabilities $P_M(A_0)$ for a continuum percolation model at
criticality in this geometry.  This is equivalent to assuming that
these probabilities are conformally-invariant properties of critical
percolation.

While the statements of universality and conformal invariance of
critical percolation do not have rigorous foundations, they are
supported by a great deal of numerical
evidence\cite{isichenko,staufferbook}. Of particular relevance to the
present work are numerical studies that have shown the probabilities
for {\it any number} of clusters of one kind to connect between the
free edges of rectangles\cite{langlands} and
cylinders\cite{huuniversal,hovi} to be universal quantities.
Furthermore, it has been shown that these probabilities can be
expressed as conformally-invariant properties of the standard
field-theoretical formulation of percolation\cite{cardy}, and this
conformal-invariance has recently been demonstrated
numerically\cite{langlandsconformal,saintaubin}. It has also been
demonstrated that the probabilities for $n$ clusters of one kind to
span the free edges of cylinders are universal
quantities\cite{huuniversal}. Since the probabilities $P_M(A_0)$ and
the spanning probabilities studied in
Refs.~\onlinecite{langlands,huuniversal,hovi,cardy,langlandsconformal,saintaubin}
have similar definitions in terms of the properties of large-scale
clusters, we believe that these works provide strong motivation for
our hypothesis that $P_M(A_0)$ are conformally-invariant quantities.

We note for reference that conformal invariance is also a property of
the continuum conduction problem in two dimensions.  For example,
suppose that $S$ and $S'$ are two regions of annular topology which
can be mapped onto each other by a conformal mapping $\bbox{r}' =
f(\bbox{r})$. Let the local conductivity tensor at each point
$\bbox{r}$ in $S$ be identical to the conductivity tensor at its image
point $\bbox{r}'$.  Then the conductance between the inner and outer
edges of $S'$ will be identical to the conductance between the inner
and outer edges of $S$.  As a particular case of this theorem, one may
use the conformal mapping described below [Eq.~(\ref{eq:conformal})]
to show that the two-terminal conductance of a Corbino disc sample
with a constant local conductivity and with an aspect ratio $A_0\equiv
(1/2\pi)\ln(r_2/r_1)$ is equal to that of a cylinder with the same
constant local conductivity and with a length that is $A_0$ times its
circumference.

\section{Connection to Numerical Studies on a Cylinder}
\label{sec:numerical}

We now show in detail how a lattice of the form illustrated in
Fig.~\ref{fig:discretize}(b) can be constructed.  We start from a
$N_x\times N_y$ square lattice whose nodes have the Cartesian
co-ordinates $(x,y)=(n_x, n_y)$, where $n_x=0,\ldots N_x-1$ and
$n_y=0,\ldots N_y-1$.  Consider the set of points defined by the
images $\{(u,v)\}$ of each node $\{(x,y)\}$ under the transformation
\begin{equation}
\label{eq:conformal}
w = e^{-2\pi i z/N_x} ,
\end{equation}
where $z= x+iy$ and $w= u+iv$.  The set of points $\{(u,v)\}$ lie in a
region the shape of a Corbino disc with inner and outer radii $r_1 =
1$, $r_2=\exp[2\pi(N_y-1)/N_x]$. The aspect ratio of the disc is
therefore $A_0 = N_y/N_x$ when $N_y\gg 1$ which we assume to be the
case.  The resulting lattice is locally square (in the limit $N_x,
N_y\gg 1$), as is guaranteed by the properties of analytic functions
in the complex plane\cite{cardyreview}, and therefore represents a
suitable discretization of critical percolation (within the
assumptions of conformal invariance).  Thus, through the use of the
above transformation, a model of bond-percolation defined on the
lattice $\{(u,v)\}$ covering a Corbino disc with aspect ratio $A_0 =
N_y/N_x$ may be related to a model of bond-percolation on a square
lattice $\{(x,y)\}$ of size $N_x\times N_y$.  To reproduce the
topology of the Corbino disc, periodic boundary conditions must be
imposed on the square lattice and a row of bonds introduced between
the edges at $n_x=0$ and $n_x=N_x-1$: this therefore represents a
bond-percolation model on a cylinder.

\begin{figure}
\vskip-4mm
\inseps{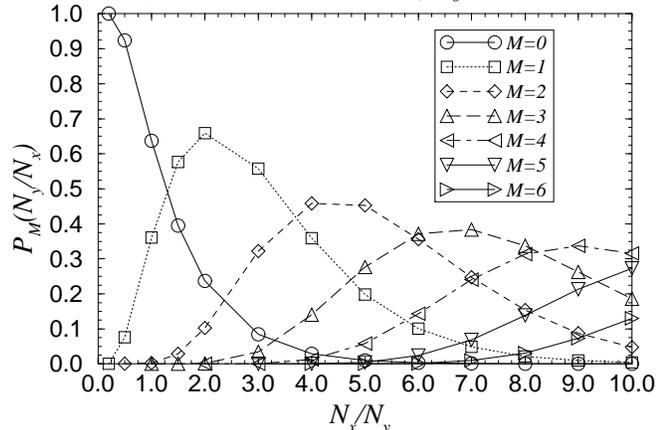}{0.5}
\caption{Probabilities for the number of alternating percolation clusters 
$M$ at the critical point of a bond-percolation model on a cylinder
with circumference $N_x$ and length $N_y$ (from the calculations
reported in Refs.~\protect\onlinecite{huhalperin}
and~\protect\onlinecite{hulin}).  By conformal invariance, the same
values describe continuum percolation on a Corbino disc with aspect
ratio $A_0=N_y/N_x$.}
\label{fig:huresults}
\end{figure}

A recent paper by two of the present authors\cite{huhalperin} contains
a discussion of bond-percolation in the cylindrical space discretized
by the $N_x\times N_y$ square lattice described above.  Numerical
calculations of the probability distribution of the number of
alternating percolation clusters are reported (the definition of this
number in terms of the distribution of clusters on a cylinder is
analogous to that on a Corbino disc, with the ends of the cylinder
playing the role of the contacts of the Corbino disc).  At the
critical point, these probabilities are found to depend only on the
ratio $N_x/N_y$ in the large-system limit ($N_x, N_y\gg 1$).  The
resulting values in the large system limit are presented in
Fig.~\ref{fig:huresults} as a function of $N_x/N_y$.

In view of the discussion of Sec.~\ref{sec:universality} and the use
of the conformal transformation~(\ref{eq:conformal}), we claim that,
at the critical point, the probability distribution of the number of
alternating percolation clusters found numerically for
bond-percolation on a cylinder with aspect ratio $N_x/N_y$ is equal to
the corresponding quantity $P_M(A_0)$ for continuum percolation on a
Corbino disc with aspect ratio $A_0=N_y/N_x$ (in the limit $N_x,
N_y\gg 1$).  The equivalence between the quantities $P_M(A_0)$ and the
results of Ref.~\onlinecite{huhalperin} is made explicit by the
relation $P_M(A_0)=F^a_M(L_1/L_2=1/A_0,0)$, where $F^a_M(L_1/L_2,x)$
is the scaling function discussed in Ref.~\onlinecite{huhalperin}
(note that the aspect ratio $R=L_1/L_2$ defined in
Refs.~\onlinecite{huhalperin,huuniversal} is the inverse of the aspect
ratio we have defined $A_0=N_y/N_x$).

To conclude our discussion of the numerical studies reported in
Ref.~\onlinecite{huhalperin}, we note that this work also presents
results for the number of alternating percolation clusters away from
the critical point.  The corresponding probability distributions are
found to satisfy a scaling form, with only the ratio of the
correlation length to the system size appearing as an additional
variable.  These results cannot be directly transformed to analogous
quantities for the Corbino disc, since under the
transformation~(\ref{eq:conformal}) the local value of the correlation
length will vary with position (due to the fact that the local lattice
constant varies with position).  To recover a suitable model for
percolation on the Corbino disc (in which the local correlation length
is uniform), one should study a cylindrical system in which the
deviation of the local bonding probability $p(\bbox{r})$ from the
critical value $p_c$ varies as $|p(\bbox{r})-p_c|^\nu\propto \exp(2\pi
y/N_x)$, where $y$ is the distance of the center of the bond along the
cylinder and $\nu=4/3$ is the correlation length
exponent\cite{isichenko}.

\section{Relation to Experimental Measurements}
\label{sec:experimental}

Combining the results shown in Fig.~\ref{fig:huresults} with
Eq.~(\ref{eq:lowtg}) leads to the probability distribution for the
low-temperature dissipative conductance of a Corbino disc under the
conditions for which our theory applies: for a sample containing
homogeneous and isotropic density fluctuations with a correlation
length that is small compared to the sample size, and within the
``critical region'' of the transition.  Since the maximum value of $M$
could occur outside of the critical region, these results place a
lower bound on the maximum peak height.  Work is under way to follow
the evolution of the number of spanning clusters as the system sweeps
through the percolation threshold\cite{hunoncritical}; this will allow
the distribution of peak-shapes (and therefore maximum peak heights)
to be determined.

Before we compare our results with experimental observations, we will
discuss how they would be affected if the samples were to contain an
additional {\it uniform} density gradient.

\begin{figure}
\inseps{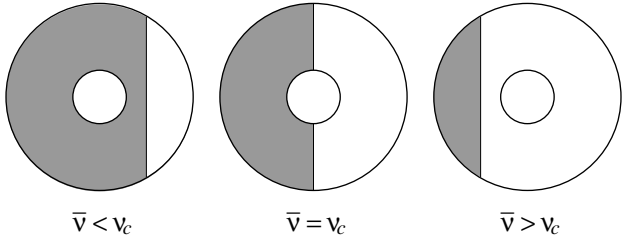}{0.28}
\vskip0.2cm
\caption{Schematic diagram of the evolution of
the distribution of local conductivity for a Corbino disc with a
uniform density gradient. During the transition between quantized Hall
plateaus, the distribution passes through a configuration with $M=1$.}
\label{fig:uniform}
\end{figure}

Consider a sample that contains a weak uniform gradient in density
(and no random component). In this case, the evolution of $M$ as a
function of the average filling fraction $\bar{\nu}$ is very simple.
The evolution of the spatial distribution of the local conductivity is
illustrated in Fig.~\ref{fig:uniform}.

For small $\bar{\nu}$ the Hall conductivity at all points in the
sample is quantized at $\sigma_1$.  As $\bar{\nu}$ increases, a
$\sigma_2$-region appears on one side of the sample and the (straight)
boundary separating the $\sigma_1$- and $\sigma_2$-regions sweeps
across the sample. Over the range of filling fraction for which this
boundary touches the inner contact, the two contacts are connected by
one $\sigma_1$-cluster and one $\sigma_2$-cluster so $M=1$.  Once the
boundary has passed the inner contact, the contacts are connected by a
single $\sigma_2$-region and $M=0$ again.  Thus, during the plateau
transition $M$ passes through a single peak with $M^{max}=1$.  This is
true for any value of the aspect ratio of the Corbino disc.

For samples drawn from an ensemble characterized by both a uniform
density gradient and homogeneous and isotropic inhomogeneities, the
probability for observing $M^{max}=1$ is increased above the value one
would calculate in the absence of the uniform gradient.  If the
uniform gradient is sufficiently large, then the probability for
$M^{max}=1$ becomes $100\%$.  To determine the threshold value of the
uniform gradient above which the random fluctuations may be neglected,
consider a sample of characteristic size $L$ containing a uniform
density gradient of size $\nabla\nu$, and isotropic and homogeneous
fluctuations with amplitude $\delta\nu$ and length scale $R_c\ll L$.
The uniform gradient introduces a spatial variation in the deviation
of the local percolation model from criticality.  Within the
transition region between quantum Hall plateaus, the local model will
be critical along a straight line passing through the sample in a
direction perpendicular to $\nabla\nu$.  However, at a distance $L$
from this line there will be a fractional deviation from criticality
of order $L\nabla\nu/\delta\nu$, so the local value of the correlation
length will be reduced to a size $\xi\simeq R_c
(\delta\nu/L\nabla\nu)^{4/3}$.  If this local correlation length is
small compared to the sample size, then, on each side of the line at
which the local model is critical, the conductivity distribution
contains a single large cluster (on one side $\sigma_1$, on the other
$\sigma_2$) and the form of the distribution is the same as in the
absence of the random fluctuations.  The condition under which the
uniform density gradient dominates and the random fluctuations can be
neglected is therefore $R_c (\delta\nu/L\nabla\nu)^{4/3}\ll L$.  One
may rewrite this condition in the form
\begin{equation}
\Delta\nu \equiv L\nabla\nu \gg \delta\nu (R_c/L)^{3/4},
\end{equation}
which shows that, for small $R_c/L$, the total change in the density
across the sample caused by the uniform gradient $\Delta\nu$ need only
be large compared to a {\it small} multiple of the typical random
density fluctuation $\delta\nu$. Thus even a small uniform gradient in
the density is sufficient to be the dominant density inhomogeneity and
cause $M^{max}=1$ for almost all samples in the ensemble.

We will now we compare our results with the measurements on the two
samples studied in Ref.~\onlinecite{rokhinsongoldman}.  It was shown
in Ref.~\onlinecite{rch} that the low-temperature conductance peak
heights for these two samples are consistent with Eq.~(\ref{eq:lowtg})
if both samples have $M^{max}=1$ (this consistency holds to an
accuracy of approximately $10\%$). Let us calculate the probability
for this to occur if the samples were drawn at random from the
ensemble of samples discussed in previous sections (characterized by
homogeneous and isotropic density fluctuations with short-range
correlations).  The probability that $M^{max}=1$ may be expressed as
the probability that $M=1$ in the critical region times the
conditional probability that in this case a larger value of $M$ does
not appear outside this region, plus the probability that $M=0$ in the
critical region times the conditional probability that in this case a
value of $M=1$ (but no larger) occurs outside of the critical region.
The conditional probabilities must be less than or equal to unity,
leading to the inequality that the probability for $M^{max}=1$ in a
Corbino disc with aspect ratio $A_0$ is less than or equal to
$P_{1}(A_0) + P_{0}(A_0)$.  The aspect ratios of the samples studied
in Ref.~\onlinecite{rokhinsongoldman} are $0.21$ and $0.32$, so the
above inequalities together with the values for $P_M(A_0)$ (estimated
by linear interpolation of the results shown in
Fig.~\ref{fig:huresults}) lead to the conclusion that the probability
for $M^{max}=1$ in these two samples is less than or equal to $0.25$
and $0.61$, respectively.  Thus the combined probability for both
samples to have $M^{max}=1$ is less than or equal to $0.15$.

The availability of only two experimental data points prevents strong
conclusions to be drawn from the comparison of our results with
experiments.  However, the reasonably small chance ($\leq 15\%$) for
both samples to exhibit $M^{max}=1$ if drawn at random from the
ensemble of samples we have assumed suggests a discrepancy between
these observations and our model.  We suggest that this small
probability may indicate the presence of long-range density
inhomogeneities in these samples.  As shown above, even a weak uniform
density gradient is sufficient to significantly increase the
probability of $M^{max}=1$.  Clearly further experiments are required
to fully determine the statistical properties of the low-temperature
non-universal conductance as a function of the aspect ratio of the
Corbino disc, both to test the theory of Ref.~\onlinecite{rch} and to
determine the form of the density inhomogeneities.  To this end, it
would be particularly interesting to perform transport measurements on
Corbino disc samples with split outer electrodes (provided the
fabrication of such a structure does not introduce significant
additional density inhomogeneities).  If the voltage of each outer
contact is held at the same value but the current through each is
measured separately, then the theory of Ref.~\onlinecite{rch} predicts
that, at low temperatures, the conductance of each contact will be
$M_i |\sigma_1-\sigma_2|$, where $M_i$ is a non-negative integer
associated with the contact $i$. For instance, if the samples studied
in Ref.~\onlinecite{rokhinsongoldman} for which $M=\sum M_i = 1$ had
split outer contacts, this prediction means that $M_i=1$ for only one
outer contact, while $M_i=0$ for all others. The statistical
properties of the conductances $M_i$ measured for a large number of
samples would provide valuable information on the form of the density
inhomogeneities.

\section{Summary}
\label{sec:summary}

An explanation for the observed ``non-universal scaling'' of the
conductance (and conductivity) peak-heights in two-dimensional
electron systems in the quantized Hall regime has recently been
proposed\cite{rch}. Within this theory, the heights of the conductance
peaks are related to the number of alternating percolation clusters
for a percolation model defined on the spatially-varying local filling
fraction in the sample.  Motivated by this work, we have studied the
statistical properties of the number of alternating percolation
clusters for Corbino disc samples. We considered the samples to
contain random density fluctuations that are isotropic and homogeneous
and have a correlation length that is small compared to the sample
size.  By appealing to conformal invariance of critical percolation,
we have argued that in the ``critical region'' of the quantum Hall
transition the probability distribution of the number of alternating
percolation clusters can be found from numerical calculations of the
same quantity for a bond-percolation model defined on a
cylinder\cite{huhalperin}. We used this identification and the
results of Ref.~\onlinecite{huhalperin} to obtain a lower bound to the
low-temperature conductance peak heights, and compared our results
with recent experimental measurements\cite{rokhinsongoldman}. The
experimental observations clearly show non-universal scaling for two
Corbino disc samples, with prefactors that are consistent with the
theory of Ref.~\onlinecite{rch} if $M^{max}=1$.  We concluded that,
within the assumptions we had made, the combined probability for
$M^{max}=1$ in these two samples is less than $15\%$.  We suggested
that this small probability could indicate a failure of our
assumptions concerning the form of the density inhomogeneities in
these samples, and that additional long-wavelength components could be
present.

\acknowledgements{We are grateful to L.~Rokhinson and 
V.~Goldman for helpful discussions. This work was in part supported by
the National Science Foundation of the USA under grant number DMR
94-16910, and the National Science Council of the Republic of China
(Taiwan) under grant numbers NSC 85-2112-M-001-007 Y and NSC
85-2112-M-001-045.}

\widetext


\begin{thebibliography}{10}

\bibitem{nonuniexpt}
R. Willett {\it et~al.}, Phys. Rev. Lett. {\bf 59}, 1776 (1987);
V.~J.~Goldman {\it et~al.} (unpublished) [the results are reproduced
in J.K. Jain, Adv. Phys. {\bf 41}, 105 (1992)]; see also the
discussion in Ref.\onlinecite{rch}.
 
\bibitem{Kucera}
J. Kucera and P. Streda, J. Phys. C {\bf 21},  4357  (1988).

\bibitem{huo}
Y. Huo, R.~E. Hetzel, and R.~N. Bhatt, Phys. Rev. Lett. {\bf 70},  481  (1993).

\bibitem{ruzin}
A.~M. Dykhne and I.~M. Ruzin, Phys. Rev. B {\bf 50},  2369  (1994);
I.~M. Ruzin and S. Feng, Phys. Rev. Lett. {\bf 74},  154  (1995).

\bibitem{rokhinsongoldman}
L.~P. Rokhinson, B. Su, and V.~J. Goldman, Solid State Comm. {\bf 96},
309 (1995); unpublished, 1995.

\bibitem{rch}
I.~M. Ruzin, N.~R. Cooper, and B.~I. Halperin, Phys. Rev. B {\bf 53},  1558
  (1996).

\bibitem{huhalperin}
C.-K. Hu and B.~I. Halperin, (unpublished).

\bibitem{isichenko}
M.~B. Isichenko, Rev. Mod. Phys. {\bf 64},  961  (1992).

\bibitem{staufferbook}
D. Stauffer and A. Aharony, {\em Introduction to Percolation Theory} (Taylor
  and Francis, London, 1992).

\bibitem{cardyreview}
J.~L. Cardy,  in {\em Phase Transitions and Critical Phenomena}, edited by C.
  Domb and J.~L. Lebowitz (Academic Press, London, 1987), Vol.~11, pp.\
  55--126.

\bibitem{langlands}
R. Langlands, C. Pichet, P. Pouliot, and Y. Saint-Aubin, J. Stat. Phys. {\bf
  67},  553  (1992).

\bibitem{huuniversal}
C.~K. Hu, C.~Y. Lin, and J.~A. Chen, Phys. Rev. Lett. {\bf 75}, 193
(1995); {\bf 75}, 2786 (E) (1995).

\bibitem{hovi}
J.~P. Hovi and A. Aharony, Physical Review E {\bf 53},  235  (1996).

\bibitem{cardy}
J.~L. Cardy, J. Phys. A {\bf 25},  L201  (1992).

\bibitem{langlandsconformal}
R. Langlands, P. Pouliot, and Y. Saint-Aubin, Bulletin of the American
  Mathematical Society {\bf 30},  1  (1994).

\bibitem{saintaubin}
Y. Saint-Aubin, Physica A {\bf 221},  41  (1995).

\bibitem{hulin}
C.-K. Hu and C.-Y. Lin, Phys. Rev. Lett. {\bf 77}, 8 (1996).

\bibitem{hunoncritical}
C.-K. Hu (unpublished).

\end{thebibliography}
\end{document}